\documentclass[onecolumn,aps,prfluids,showpacs,longbibliography]{revtex4-2}

\usepackage{tabularx}
\usepackage[latin1]{inputenc}
\usepackage{graphicx}
\usepackage{siunitx}
\usepackage{latexsym}
\usepackage{amssymb}
\usepackage{amsmath}
\usepackage{color}
\usepackage{tikz}
\usepackage{lipsum}
\usepackage{soul}
\usepackage{cancel}

\usepackage{amssymb}

\begin{document}
\def\d{{\rm d}}
\def\dipolo{{\bf d}}
\def\ex{{\rm e}}
\def\dbar{{\mathchar'26\mkern-12mu d}}
\def\tr{{\rm tr}}
\def\v{{\bf v}}
\def\sign{{\rm sign}}
\def\r{{\bf r}}
\def\l{{\bf l}}
\def\V{{\bf V}}
\def\u{{\bf u}}
\def\x{{\bf x}}
\def\A{{\bf A}}
\def\a{{\bf a}}
\def\c{{\bf c}}
\def\f{{\bf f}}
\def\e{{\bf e}}
\def\p{{\bf p}}
\def\k{{\bf k}}
\def\n{{\bf n}}
\def\m{{\bf m}}
\def\q{{\bf q}}
\def\g{{\bf g}}
\def\E{{\bf E}}
\def\D{{\bf D}}
\def\iv{{\rm iv}}
\def\B{{\bf B}}
\def\F{{\bf F}}
\def\Y{{\bf Y}}
\def\Y{{\bf Y}}
\def\y{{\bf y}}
\def\z{{\bf z}}
\def\X{{\bf X}}
\def\Y{{\bf Y}}
\def\C{{\mathcal{C}}}
\def\calE{{\cal E}}
\def\L{{\bf L}}
\def\P{{\cal P}}
\def\bxi{{\boldsymbol{\xi}}}
\def\balpha{{\boldsymbol{\alpha}}}
\def\smalze{{\scriptscriptstyle{(0)}}}
\def\smalunmez{{\scriptscriptstyle{(1/2)}}}
\def\smalun{{\scriptscriptstyle{(1)}}}
\def\smaldu{{\scriptscriptstyle{(2)}}}
\def\smaltr{{\scriptscriptstyle{(3)}}}
\def\smalqu{{\scriptscriptstyle{(4)}}}
\def\smalci{{\scriptscriptstyle{(5)}}}
\def\smalse{{\scriptscriptstyle{(6)}}}
\def\smaln{{\scriptscriptstyle{(n)}}}
\def\smalH{{\scriptscriptstyle{H}}}
\def\smalNH{{\scriptscriptstyle{NH}}}
\def\smalL{{\scriptscriptstyle{L}}}
\def\smalint{{\scriptscriptstyle{int}}}
\def\PDF{{\mathfrak f}}
\def\bPsi{{\boldsymbol{\Psi}}}
\def\bOmega{{\boldsymbol{\Omega}}}
\def\boldeta{{\boldsymbol{\eta}}}
\def\J{{\bf J}}
\def\albedo{{\cal A}}
\def\latent{{\cal L}}
\def\im{{\rm i}}
\def\smalf{{\scriptscriptstyle{f}}}
\def\smali{{\scriptscriptstyle{i}}}
\def\smalS{{\scriptscriptstyle{(S)}}}
\def\Pness{P_{\scriptscriptstyle{NESS}}}
\def\beq{\begin{eqnarray}}
\def\eeq{\end{eqnarray}}



\title{Zero-gravity convection in a closed duct as the realization of a 
thermal machine}
\author{Piero Olla}
\thanks{Email address for correspondence: olla@dsf.unica.it}
\affiliation{ISAC-CNR and INFN, Sez. Cagliari, I--09042 Monserrato, Italy.}

\begin{abstract}
The possibility of convection in a wall-heated unstirred simple fluid in zero-gravity 
conditions is discussed. It is shown that a low-Prandtl-number fluid 
possesses a stable circulating state in zero gravity, without destabilization of the diffusive 
state, provided geometric inhomogeneity allows the thermal cycle to perform net mechanical work.
The analysis provides a quantitative explanation of why purely volumetric effects
cannot induce by themselves convection in conventional fluids under zero-gravity conditions.


\end{abstract}
\maketitle
Buoyancy-induced convection is absent by definition in zero-gravity (zero-$g$) environments.
Much less effective forms of heat transfer, primarily diffusion, take the stage,
making ventilation crucial for the thermal management aboard spacecraft \cite{broyan10}.

Actually, non-diffusive zero-$g$ heat transport is still possible
at the interface between different fluids (or different phases of the same fluid),
because of the Marangoni effect \cite{kawamura}. Convection becomes possible again if the fluid is
vibrated as a whole---a phenomenon called thermal vibrational convection 
\cite{beysens06,mialdun08,huang24}.
Thermoacoustic coupling between temperature gradients and acoustic modes 
\cite{swift}, as well as thermo-osmotic flows driven by temperature gradients along 
solid-fluid interfaces \cite{anzini22,bregulla16},
may also lead to fluid motion.
Zero-$g$ convection becomes a possibility also
near the critical point, where the 
thermal expansion coefficient diverges, and viscosity simultaneously vanishes 
\cite{beysens11,zappoli03}.
Localized motions can be observed
if the fluid is heated internally, say by a laser \cite{bar07,kostoglou11}.
In contrast, experiments with heating and cooling at solid boundaries show predominantly diffusive transport under microgravity conditions 
\cite{colombani00}. 
A similar situation is observed in 
granular gases \cite{rodriguez20}.

A simple explanation of such a state of affairs is that direct volumetric effects are
too small to counteract the viscosity-induced slow-down of plumes near the walls.
However, a more quantitative description of the process is  desirable.
Also, the absence of a linear instability does not by itself exclude 
finite-amplitude convective states, where the volumetric contribution, in certain
parameter ranges, could become substantial. A general investigation of the possibility
of stationary currents in zero-gravity heated fluids is therefore of interest.

\begin{figure}[h]
\begin{center}
\includegraphics[draft=false,width=9cm]{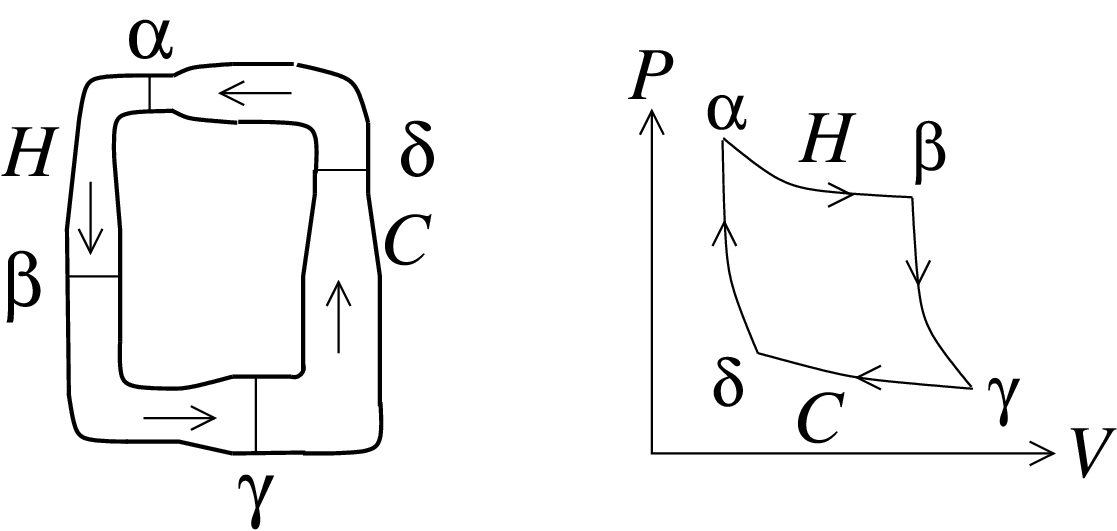}
\caption{The duct geometry. On the right, sketch of the thermal cycle of a fluid parcel
in the duct.
}
\label{dipfig1}
\end{center}
\end{figure}

The simplest geometry in which to investigate such phenomena is the closed duct 
in Fig. \ref{dipfig1}.
The idea is that the flow of a compressible fluid could support itself through conversion
of part of the heat received by the hot reservoir to kinetic energy. As in the case of 
a thermal machine, the process would be induced by expansion in the sections of the duct where
the radius increases.  Work against the friction 
forces converts the kinetic energy back into heat, which is then delivered to the cold reservoir
in what amounts essentially to a thermal cycle, as illustrated in the figure. 

The aspect ratio of the
duct is assumed sufficiently large to allow a one-dimensional description of the flow, which
could represent a flow tube in a larger configuration, such as a hypothetical
convective thermal dipole realized by a couple of localized heat sources and sinks, 
or a physical duct, of the kind studied in the thermosyphon literature \cite{welander67,fichera03}.
Sections $\alpha\beta$ (hot reservoir, $H$) and $\gamma\delta$ (cold reservoir, $C$) in the duct 
are thermostatted; sections $\beta\gamma$ and
$\delta\alpha$ are thermoisolated (of course, no thermoisolated sections would be present 
in the case of a flow tube).

To fix the ideas, let us assume equal lengths for the different sections of the tube,
$L_{\alpha\beta}=L_{\beta\gamma}=L_{\gamma\delta}=L_{\delta\alpha}=L$, and a piecewise
linear profile of the duct radius $R(x)$.
Define
\beq
\hat R=R/\langle R^2\rangle^{1/2},
\eeq
where $\langle.\rangle$ indicates average along the duct, 
$\int\d^2x_\perp$ is the integral
over a transversal section of the tube, and introduce the integrated density and
current
\beq
\rho(x,t)=\frac{1}{\langle R^2\rangle}\int\d^2 x_\perp\ \hat\rho(\x,t),
\qquad
J(x,t)=\frac{1}{\langle R^2\rangle}\int\d^2 x_\perp\ \hat J(\x,t),
\eeq
the latter defined positive in the direction of the arrows in figure. 
We posit that the radial dependence of $\hat J$ and $\hat\rho$ can be disregarded for $R\ll L$.
Thus, $\hat\rho(\x,t)=\rho(x,t)/\hat R^2(x)$,
and the fluid velocity
\beq
u=J/\rho
\label{J}
\eeq
is defined unambiguously. Finally, we assume that $\hat R$ is linear, separately, 
in each section of the duct.
 
We want to investigate whether the system can sustain a constant current. To study the
problem, we must convert the conservation equations that describe the dynamics to a 
one-dimensional form. 
The mass and total energy conservation equations read
\beq
&&\partial_t\rho+\partial_xJ=0,
\label{Cmass}
\\
&&\partial_t[\rho(u^2+2c_VT)]+\partial_x[J(u^2+2c_PT)]=2q
+2\kappa\partial_x^2T,
\label{Cenergy}
\eeq
where $T$ is the temperature, 
and $q$ is the thermostat contribution to the heat budget.
In Eq. (\ref{Cenergy}) we have set the Boltzmann constant equal to the 
molecular mass, in such a way that $T$ has
dimensions of velocity squared, and we are assuming an ideal gas law of state,
$P=\hat\rho T$, $c_P=c_V+1$. We verify that $u^2/2$ and $c_VT$ are the kinetic and
thermal energy per unit mass, and $\langle R^2\rangle JT=uR^2P$ is the power delivered by the
pressure force at the given section of the duct.
We consider a linear law for the heat transfer from the thermostats to the fluid,
\beq
q=sc_P (T_{th}-T),
\qquad
s=\begin{cases}
\sigma, & x\in[x_\alpha,x_\beta]\cup[x_\gamma,x_\delta],
\\
0,\ &{\rm otherwise},
\end{cases}
\label{conduction}
\eeq
and for simplicity assume in the thermostats a piecewise temperature profile,
$T_{th}(x)=T_{H,C}$.

To close the system, we need one more equation, which we can obtain from thermodynamics.
The entropy $S_L$ of a Lagrangian fluid element $V_L$ containing $N$ molecules obeys
\beq
k_BT\dot S_L=\hat\rho T\dot V_L+Nk_Bc_V\dot T_L
=Nk_B\Big(-\frac{\dot{\hat\rho}_L}{\hat\rho_L}+c_V\frac{\dot T_L}{T_L}\Big)
=(V_L/\hat R^2)
[\gamma u^2+\kappa\partial_x^2T+q],
\label{thermodynamics}
\eeq
where the terms in square brackets on the RHS of the equation are, in the order, the
heating contribution from mechanical friction---assumed linear in  $u$, 
heat diffusion, and heat transfer to and from
the thermostats. From 
Eq. (\ref{thermodynamics}) we obtain the equation for the entropy budget,
\beq
[\rho\partial_t+J\partial_x]\ln(T^{c_V}/\hat\rho)=(\gamma u^2+q)/T
+\kappa[\partial_x(\partial_xT/T)+(\partial_xT/T)^2].
\label{Centropy}
\eeq
We can obtain an alternative description of the
dynamics, with a heat transport equation or a momentum conservation law replacing 
one of the former equations. From
Eq. (\ref{Centropy}), we readily obtain the heat transport equation
\beq
c_V\rho(\partial_t+u\partial_x)T+\hat\rho T\partial_x(\hat R^2u)=
\gamma u^2+\kappa\partial_x^2T+q,
\label{heat_transport}
\eeq
which, multiplied by two and subtracted
from Eq. (\ref{Cenergy}), gives us
the balance equation for the kinetic energy
\beq
\partial_t(\rho u^2)+\partial_x[J(u^2+2T)]-2\hat\rho T\partial_x(\hat R^2u)=-2\gamma u^2,
\label{K budget}
\eeq
consistent with the momentum conservation law
\beq
&&\partial_tJ+\partial_x(uJ)+\hat R^2\partial_xP=-\gamma u.
\label{Cmomentum}
\eeq
We integrate Eq. (\ref{Cmomentum}) around the duct, reaching the important conclusion 
that no stationary finite-current regime can be realized if $\partial_x\hat R= 0$.

Let us take units such that
\beq
L=\langle\rho\rangle=
(T_H+T_C)/2=1,
\eeq
and write
\beq
T_{H,C}=1\pm T_H^\smalun\epsilon\quad {\rm and}\quad
\hat R=1+\hat R^\smalun\epsilon.
\eeq
with $\epsilon\ll 1$. We assume a regular expansion in $\epsilon$ for all quantities
in the problem $Y=\sum_n Y^\smaln\epsilon^n$, except $R$, which only enters 
Eqs. (\ref{Cmass}-\ref{Cmomentum}) through the ratio $\hat R$. We consider first
a low Mach number regime ${\rm Ma}\simeq J/c_s=O(\epsilon)$, where $c_s=\sqrt{c_P/c_V}$ is 
the sound speed.

A finite-current regime requires that dissipation does not dominate over the kinetic
terms in Eqs. (\ref{Cenergy}-\ref{Cmomentum}). Let us assume stationarity, and
focus on the entropy production term in the bulk of the fluid
\footnote{Entropy
is produced in the loop by the diffusive heat transfer from the reservoirs to
the fluid in the thermostatted sections of the duct, which is offset by the entropy increase
in the thermostats. 
},
\beq
\langle\dot S\rangle=\langle[(\gamma J^2+q)/T+\kappa(\partial_xT/T)^2]\rangle=0.
\label{dot S}
\eeq 
To obtain $\langle q/T\rangle$ we need $T$, which we 
obtain from Eq.  (\ref{Cenergy}).
To lowest order in $\epsilon$, 
\beq
J^\smalun\partial_xT^\smalun=s^\smalun
(T_{th}^\smalun-T^\smalun),
\label{Cener0}
\eeq
which has the solution
\beq
\begin{cases}
T^\smalun(x\in[x_\alpha,x_\beta])=\{1-2\exp[-k(x-x_\alpha)]/[1+\exp(-k)]\}T_H^\smalun,
\\
T^\smalun(x\in[x_\beta,x_\gamma])=[1-\exp(-k)]/[1+\exp(-k)]T_H^\smalun,
\\
T^\smalun(x\in[x_\gamma,x_\delta])=\{-1+2\exp[-k(x-x_\gamma)]/[1+\exp(-k)]\}T_H^\smalun,
\\
T^\smalun(x\in[x_\delta,x_\alpha])=-[1-\exp(- k)]/[1+\exp(-k)]T_H^\smalun,
\end{cases}
\label{Tsmalun}
\eeq
where $k=\sigma^\smalun/J^\smalun$. However, from Eq. (\ref{Cenergy}),
\beq
\left\langle\frac{q^\smaldu}{1+\epsilon T^\smalun}\right\rangle&\propto&
\left\langle\frac{\epsilon\partial_xT^\smalun}{1+\epsilon T^\smalun}\right\rangle_{\alpha\beta}
+\left\langle\frac{\epsilon\partial_xT^\smalun}{1+\epsilon T^\smalun}\right\rangle_{\gamma\delta}
\nonumber
\\
&\propto&\int_{x_\alpha}^{x_\beta}\d\ln(1+\epsilon T^\smalun(x))
+\int_{x_\gamma}^{x_\delta}
\d\ln(1+\epsilon T^\smalun(x))=0,
\eeq 
which tells us that to 
determine the heat flux contribution to entropy production, we must pursue the
expansion of $T$ to higher order in $\epsilon$.
From inspection of Eq. (\ref{Cenergy}), 
the next contribution to $T$ is $O(\epsilon^3)$,
hence $\langle q/T\rangle=O(\epsilon^5)$, which for $J=O(\epsilon)$ gives us 
the condition 
\beq
\gamma=O(\epsilon^3),\qquad\kappa=O(\epsilon^3).
\label{gamma kappa}
\eeq
Equation (\ref{gamma kappa})
provides a quantitative content to the statement that the mechanism preventing the formation of
plumes near solid surfaces in zero-$g$ conditions is viscosity.
Indeed, suppose conduction is the mechanism governing the heat transfer from the reservoirs to
the fluid. In a laminar regime, we would have
$\sigma\sim\kappa R^{-2}$ and
$\gamma\sim\nu R^{-2}$, where $\nu$ is the kinematic viscosity.
The two conditions $\sigma=O(\epsilon)$ and $\gamma=O(\epsilon^3)$ then require
${\rm Pr}=O(\epsilon^2)$.
Turbulence does not improve things, as ${\rm Pr}_{turb}\sim 1\Rightarrow
\gamma_{turb}/\sigma_{turb} =O(1)$.
We must assume some alternative
means of heat transfer, such as e.g. radiation, or consider a different geometry,
such as that of a flow tube with immaterial boundaries.
In the latter case, mechanical dissipation would arise from gradients at scale $L=1$,
$\gamma J^2\sim \nu\langle(\partial_{x_\perp}J)^2\rangle\sim \nu J^2/L^2\sim \nu J^2$,
to be compared with $\kappa\langle (\partial_xT)^2\rangle$.

With all the caveats in place, we can proceed to evaluate the stationary current. 
We make the ansatz (to be verified a posteriori)
that for small Ma, close to the fixed point, one can approximate $\partial_xJ=0$.
We then rewrite (\ref{Cmomentum}) as an equation for the pressure, which
reads, for $\partial_xJ=0$,
\beq
\partial_xP+A(P)=-\dot JB(P),
\qquad
P=\frac{T\rho}{\hat R^2},
\label{Cmomentum1}
\eeq
where 
\beq
A=BJ\Big[\frac{\gamma T}{P\hat R^2}+\frac{J}{P}\partial_x\frac{T}{\hat R^2}\Big],
\qquad
B=\frac{1}{\hat R^2-J^2T/(\hat RP)^2},
\label{AB}
\eeq
and periodicity of $P$ imposes
\beq
\dot J=-\langle A\rangle/\langle B\rangle.
\label{Cmomentum2}
\eeq

To obtain the pressure,
we need the density. From 
Eq. (\ref{Centropy}) we get, at stationarity,
\beq  
J^\smalun\partial_x(c_VT^\smalun+2\hat R^\smalun-\rho^\smalun)=c_PJ^\smalun\partial_xT^\smalun
\Rightarrow
\rho^\smalun=-T^\smalun+2\hat R^\smalun,
\label{rhosmalun} 
\eeq 
which tells us that $P^\smalun=0$.
We substitute Eqs. (\ref{Tsmalun}) and (\ref{rhosmalun}) into Eq. (\ref{Cmomentum2}), and obtain
\beq
&&\dot J^\smalqu=
-[\gamma^\smaltr-2J^\smalun\langle R^\smalun\partial_xT^\smalun\rangle]J^\smalun=
-[\gamma^\smaltr+2J^\smalun\langle T^\smalun\partial_xR^\smalun\rangle]J^\smalun
\nonumber
\\
&&=-\left\{\gamma^\smaltr +\left[\frac{\hat R_{\beta\alpha}-\hat R_{\delta\gamma}}{2}
\int_{x_\alpha}^{x_\beta}\d x\ T^\smalun(x)+
\frac{\hat R_{\gamma\beta}-\hat R_{\alpha\delta}}{2}\int_{x_\beta}^{x_\gamma}\d x\ T^\smalun(x)
\right]J^\smalun\right\}J^\smalun
\nonumber
\\
&&=-\Big[1+(aF_a+bF_b)J^\smalun\Big]
\gamma^\smaltr J^\smalun:=F(J^\smalun),
\label{dJdtqu}
\eeq
where 
\beq
a=\frac{T_H^\smalun}{2\gamma^\smaltr}(\hat R^\smalun_{\gamma\beta}-\hat R^\smalun_{\alpha\delta}),
\quad
b=\frac{T_H^\smalun}{2\gamma^\smaltr}( \hat R^\smalun_{\beta\alpha}-\hat R^\smalun_{\delta\gamma}),
\quad F_a=\frac{1-\ex^{-k}}{1+\ex^{-k}},
\quad F_b=1-\frac{2}{k}F_a,
\label{ab} 
\eeq
and we have exploited $P^\smalun=0$.

We can exploit Eq. (\ref{dJdtqu}) to study the approach to the fixed point, however, we must
first check the validity of the ansatz $\partial_xJ=0$.
Mass conservation dictates
$\langle\partial_t\rho\rangle=0$, which means that the lowest order contribution to $C$ 
involves a product of $J^\smalun\partial_t\rho^\smalun$ and another $O(\epsilon)$ term. The 
finite-compressibility correction to $\partial_t$ is thus $O(\epsilon^4)$. On the other hand,
Eq. (\ref{dJdtqu}) gives us $\partial_t=O(\epsilon^3)$. We can thus
safely use Eq. (\ref{dJdtqu}) to determine the stability properties of the fixed points.

\begin{figure}[h]
\begin{center}
\includegraphics[draft=false,scale=0.7]{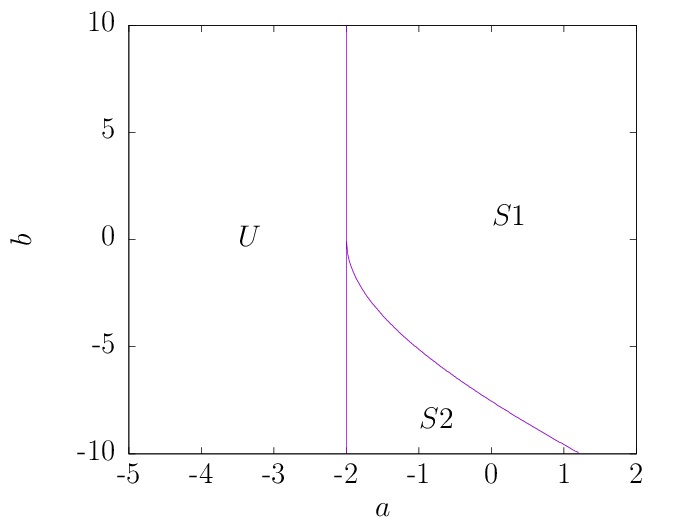}
\caption{Stability portrait of the system in the perturbative range.
}
\label{F5fig}
\end{center}
\end{figure}

At small $J$, $F\simeq -\gamma^\smaltr J^\smalun$, which tells us that the diffusive fixed 
point $\bar J^\smalun=0$ is always linearly stable. At large $J^\smalun$, instead, 
$F\simeq \gamma^\smaltr(J^\smalun)^2
(1+b/2)$, which tells us that for $b<-2$ an unstable fixed point must occur at $J>0$.
The stability diagram in Fig. \ref{F5fig} 
confirms the results, and points to the existence of two more regimes,
one of global stability for the diffusive fixed point ($S1$), one of existence of 
a stable fixed point at $\bar J_s$ paired with an unstable fixed point at $\bar J_u$,
$0<\bar J_u<\bar J_s$.

We can use an iterative procedure to solve Eqs. (\ref{Cenergy}) and (\ref{Cmomentum1}) in
the general case (numerical code available in the Supplemental Material \cite{supp})
At each step we solve Eq. (\ref{Cmomentum}) using the triplet $(\rho_n,J_n,T_n)$. 
To this aim we operate an inner loop
\beq
\partial_xP_{n,{l+1}}=\frac{B(P_{n,l},T_n,J_n)\langle A(P_{n,l},T_n,J_n)
\rangle}{\langle B(P_{n,l},T_n,J_n)\rangle}
-A(P_{n,l},T_n,J_n) :=H_n,
\label{Cmomentum3}
\eeq
where $P_{n,0}=P_n=T_n\rho_n/\hat R^2$. The solution of Eq. (\ref{Cmomentum3}) is in the form
$P_{n,l+1}(x)=P_{n,l+1}(0)+\int_0^x\d y\ H_n(y)$, where the constant $P_{n,l+1}(0)$ is fixed by
mass conservation, $\langle P_{n,l+1}\hat R^2/T_n\rangle=1$. At convergence, $l=l_*$, 
the solution of Eq. (\ref{Cmomentum3}) gives us $P_{n+1}=P_{n,l_*}$, which closes the inner loop. 
We then define 
$\rho_{n+1}=\hat R^2_{n+1}/T_n$, and use the value of $\dot J$ obtained by solving
Eq. (\ref{Cmomentum2}) with $T=T_n$ and $P=P_{n+1}$, to update the current.
The new density $\rho_{n+1}$ and the new current $J_{n+1}$ are then used in Eq. (\ref{Cenergy})
to obtain $T_{n+1}$.

\begin{figure}[h]
\begin{center}
\includegraphics[draft=false,scale=1.1]{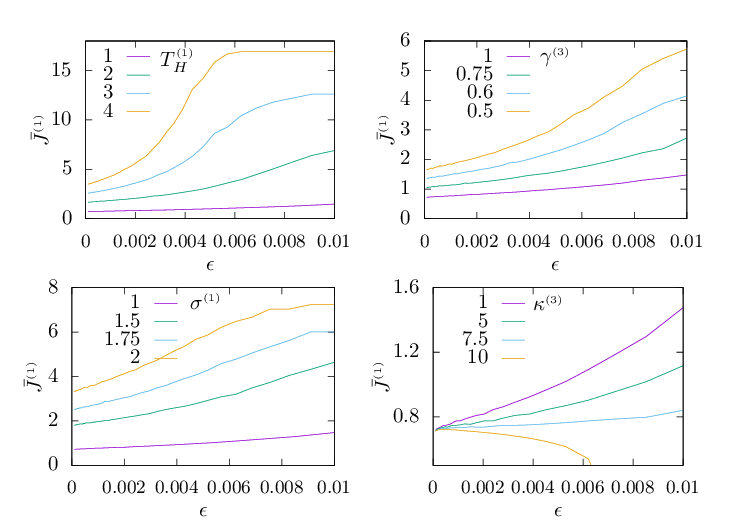}
\caption{Stationary current profiles in perturbative regime $S2$.
Values of the parameters not in legend:
$T^\smalun_H=\sigma^\smalun=\gamma^\smaltr=\kappa^\smaltr=1$.
Duct radii: $R^\smalun_\alpha=-0.25$,
$R_\beta^\smalun=0.75$, $R_\gamma^\smalun=-10.75$, $R^\smalun_\delta=11.25$.
}
\label{JSfig}
\end{center}
\end{figure}

We have used the procedure to determine the deviations of $\bar J$ 
from the prediction of Eq. (\ref{dJdtqu}). The result is shown in Fig. \ref{JSfig}.
To prevent the formation of small scale disturbances in the numerical
solution of Eqs. (\ref{Cenergy}) and 
(\ref{Cmomentum3}), the profiles of $\hat R$, $s$ and $T_{th}$ have been smoothed by spline
interpolation at scale $l_w=0.1$.  Equation
(\ref{Cenergy}) has been solved by
a two-step scheme, with one step upstream for diffusion, the second downstream for all the 
other terms in the equation, with a discretization $\Delta x=1/1024$.

\begin{figure}[h]
\begin{center}
\includegraphics[draft=false,scale=0.85]{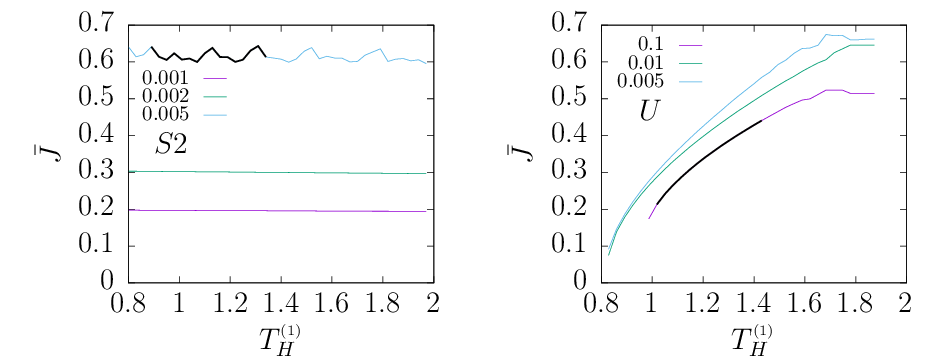}
\caption{Stationary current profiles in the non-perturbative regime. Sections of the 
curves in black indicate instability regions.  Parameters in both
cases $S2$ and $U$, $\sigma^\smalun=\gamma^\smaltr=\kappa^\smaltr=1$. Duct radii in 
case $S2$ as in Fig. \ref{JSfig}.
Duct radii in case $U$, $R_\alpha^\smalun=0.25$, $R^\smalun_\beta=2.25$, 
$R_\gamma^\smalun=R^\smalun_\delta=-1.25$.
}
\label{JNPfig}
\end{center}
\end{figure}

Using the same procedure in the non-perturbative regime reveals the presence 
of non-perturbative fixed points in regions $S2$ and $U$ of the $ab$ plane. 
A possible mechanism leading to the existence of a non-perturbative fixed point is the change of 
sign of $B$ at large $J$ [see Eq. (\ref{AB})]. However, the hypothesis is weakened
by the fact that no additional fixed points have been observed in region $S1$.
The dependence
of the stationary current on parameters $\epsilon$ and $T_H$ is shown 
in Fig. \ref{JNPfig}. 

\begin{figure}[h]
\begin{center}
\includegraphics[draft=false,scale=0.7]{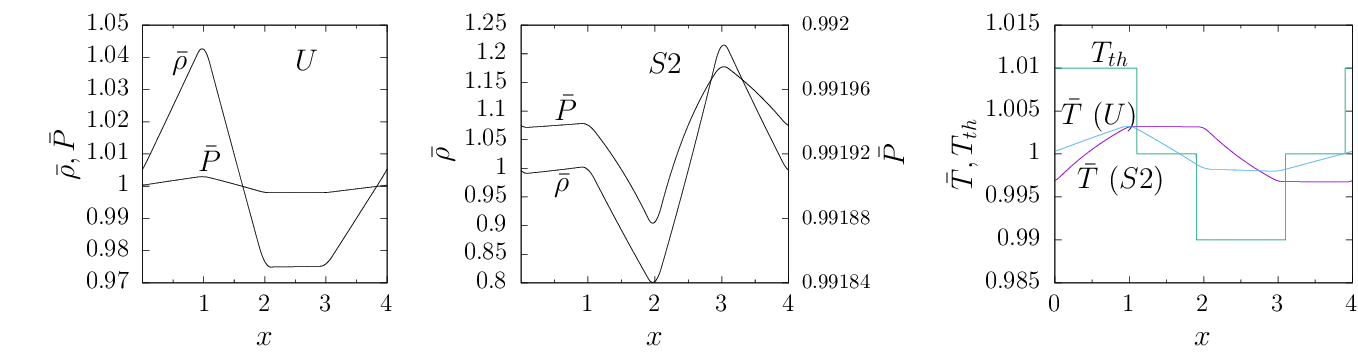}
\caption{Profiles of the density, temperature and pressure profile in the perturbative $(S2)$ 
and non-perturbative $(U)$ case for $\epsilon=0.01$ and $T^\smalun_H=1$. Values of the other
parameters as in Figs. \ref{JSfig} and \ref{JNPfig}. Note the different magnitude of the
pressure disturbance in the perturbative and non-perturbative regimes.
}
\label{rhoPTTthfig}
\end{center}
\end{figure}

It is interesting to note that, as illustrated in Fig. \ref{rhoPTTthfig},
the non-uniform component of $\rho$ and $T$
continues to be small also for $J=O(1)$, 
even though the scaling in $\epsilon$ remains unclear, suggesting
that the estimate for $q$ in Eq. (\ref{dot S}), and consequently also the condition on the 
scaling of $\gamma$ and $\kappa$ in Eq. (\ref{gamma kappa}), continue to hold in the 
non-perturbative regime. In other words, 
invoking the possibility of a non-perturbative regime thus does not
simplify the task of triggering convection in a zero-$g$ fluid heated at the walls.

We could try to use Eqs. (\ref{Cmomentum1}-\ref{Cmomentum3}) to solve the time-dependent dynamics 
in the non-perturbative regime, however, the 
incompressibility constraint that is artificially imposed at each step in the iteration procedure
Eq. (\ref{Cmomentum3}) is no longer justified. Indeed, following a fictitious $\partial_xJ=0$
trajectory
from a perturbative fixed point to its non-perturbative counterpart, would suggest that 
the non-perturbative fixed point is stable in $U$, unstable in $S2$. Instead, analysis 
of the eigenvalue spectrum near the fixed point by a dedicated software such as Dedalus shows,
as illustrated in Fig. \ref{JNPfig}, that stable and unstable
fixed points can be found in both regions $U$ and $S2$ of the parameter space 
(Dedalus script available in the Supplemental Material \cite{supp}).

The stability of the fixed points rests on the sign of the real part of the eigenvalues of
the system of equations
\beq
\partial_t\tilde X_i=\Gamma\tilde X_i=L_{ij}\tilde X_j,
\eeq
obtained by linearizing Eqs. (\ref{Cmass}), (\ref{Cmomentum}) and (\ref{heat_transport}) around
the position dependent background $\bar\X=(\bar\rho,\bar J,\bar T)$, $\tilde\X=\X-\bar\X$.
The retrieved eigenvalue spectra $\Gamma_\n=\Gamma_{r,\n}+\im
\Gamma_{i,\n}$ 
do not appear to be sensitive to the choice of basis in Dedalus (Fourier bases
of 64, 128 and 256 modes). 
Two examples of eigenvalue spectra retrieved by the software are illustrated in Table \ref{table1}.

\begin{table}
\centering
\small
\caption{Lower portion of two eigenvalue spectra for
$\epsilon=0.005$ and $T_H^\smalun=1.6$.
Other parameters as in Fig. \ref{JNPfig}. }
\setlength{\tabcolsep}{2pt}
\resizebox{0.72\textwidth}{!}{%
\begin{tabular}{
 | S S
 | S S|
}
\hline
\multicolumn{2}{|c|}{$S2$} &
\multicolumn{2}{c|}{$U$} \\
\hline
$\Gamma_r$ &$\Gamma_i$&
$\Gamma_r$ &$\Gamma_i$\\
 0.0       &  0.0      &  0.0     &  0.0      \\
-4.17e-3   &  1.16e-14 & -1.0e-3  &  3.09e-15 \\
-1.01e-3   &  8.66e-13 & -4.17e-3 &  3.99e-15 \\
-3.86e-3   &  0.996    & -3.49e-3 &  0.666    \\
-1.62e-3   &  1.022    & -6.45e-3 &  1.33     \\
-6.47e-3   &  1.99     & -1.88e-3 &  1.36     \\
-3.02e-3   &  2.05     & -1.38e-2 &  2.0      \\
-1.91e-2   &  3.0      & -1.83e-2 &  2.66     \\
 6.38e-3   &  3.01     & -1.45e-3 &  2.69     \\
-4.73e-3   &  3.07     & -2.86e-3 &  2.72     \\
-1.83e-3   &  3.98     & -2.72e-2 &  3.33     \\
\hline
\end{tabular}
}
\label{table1}
\end{table}
The modes mirror those in the constant-radius case, which for $\hat R$ constant can be 
decomposed into an orthogonal basis:
\begin{itemize}
\item
A zero-mode $\tilde\X=(const.,0,0)$. 
\item
A slow mode, $\tilde \X=(0,const.,0)$, whose dynamics in the perturbative 
$\partial_x\hat R\ne 0$ regime obeys Eq. (\ref{dJdtqu}).
\item
A slow mode $\tilde \X=(0,0,const.)$ describing the damping of constant temperature
disturbances by the thermostats. 
\item
The tower of acoustic and entropy modes
$\tilde\X_{n,\pm}(x,t)=\tilde\X(0,0)\ex^{\Gamma_rt}\cos\{k_n[x+(\bar J\pm c_s)t]\}$, 
$\tilde\X_{n,0}(x,t)=\tilde\X(0,0)\ex^{\Gamma_rt}\cos[k_n(x+\bar Jt)]$, where 
$k_n=n\pi/2$ and $c_s=\sqrt{c_P/c_V}$. 
\end{itemize}
In all the cases considered, destabilization of the 
fixed point was caused by modes near the lower end of the spectrum, even though the specific
mode was in the different cases typically not the same.

As regards the fate of trajectories originating from unstable fixed points, limit cycles and
chaos are an obvious possibility. An intriguing suggestion from the thermosyphon literature
is the existence of a low-dimensional inertial manifold, opening the way to
a description of the dynamics in terms of a restricted number of modes 
\cite{rodriguez98,fichera03}.
In the present case, however, such modes are likely to be geometry-dependent and non-orthogonal 
in nature.
\vskip 10pt

The takeaway of the analysis is that finite-current, stable stationary regimes
can be realized in a significant portion of the parameter space, provided dissipation is small. 
The condition limits the realization of the system to low Prandtl number fluids, non-contact
heating and cooling, or a flow tube embedded in a larger geometry rather than a narrow 
solid-walled duct. The present analysis thus provides a quantitative description of the
mechanism preventing volumetric effects from trigger convection in wall-heated simple
fluids in zero-gravity conditions.

If the conditions for a stationary current are satisfied, finite-current fixed points, 
each with its own basin of attraction in the $(\rho,J,T)$ space, coexist with the diffusive
fixed point, which remains stable in all the regimes considered.
The situation dramatically differs from that in a Rayleigh-Benard cell, in which the diffusive
fixed point becomes unstable at the critical value of a control parameter (the Rayleigh number).

It is interesting to note the essential role played by geometry: a thermostatted
constant-radius closed loop cannot support a finite current. The variation of the radius along
the loop is precisely what makes the system an example of a thermal machine, in which part
of the heat ceded by the hot thermostat is converted into work against the friction
forces on the fluid.

\bibliography{sample}

@PREAMBLE{
 "\providecommand{\noopsort}[1]{}"
 # "\providecommand{\singleletter}[1]{#1}%"
}

@misc{supp,
author = {},
title = {},
howpublished = "See supplemental material at \url{URL_will_be_inserted_by_publisher
} for the numerical codes utilized in the paper.",
year = {},
note = " "}

@article{huang24,
  title={Unifying constitutive law of vibroconvective turbulence in microgravity},
  author={Huang, Ze-Lin and Wu, Jian-Zhao and Guo, Xi-Li and Zhao, Chao-Ben and Wang, Bo-Fu and Chong, Kai Leong and Zhou, Quan},
  journal={Journal of Fluid Mechanics},
  volume={987},
  pages={A14},
  year={2024}
}

@article{welander67,
  title={On the oscillatory instability of a differentially heated fluid loop},
  author={Welander, Pierre},
  journal={Journal of Fluid Mechanics},
  volume={29},
  pages={17--30},
  year={1967}
}

@article{fichera03,
  title={Modelling and control of rectangular natural circulation loops},
  author={Fichera, A and Pagano, A},
  journal={International journal of Heat and Mass Transfer},
  volume={46},
  pages={2425--2444},
  year={2003}
}

@book{kawamura,
  title={Thermocapillary Convection in Microgravity: Thermohydrodynamic Experiment in Kibo Aboard International Space Station},
  author={Kawamura, Hiroshi and Nishino, Koichi and Matsumoto, Satoshi and Ueno, Ichiro and Yano, Taishi},
  year={2025},
  publisher={Springer Nature}
}

@article{mialdun08,
  title={Experimental evidence of thermal vibrational convection in a nonuniformly heated fluid in a reduced gravity environment},
  author={Mialdun, Aliaksndr and Ryzhkov, II and Melnikov, DE and Shevtsova, Valentina},
  journal={Physical Review Letters},
  volume={101},
  pages={084501},
  year={2008}
}

@article{rodriguez98,
  title={Diffusion induced chaos in a closed loop thermosyphon},
  author={Rodr{\'\i}guez-Bernal, An{\'\i}bal and Van Vleck, Erik S},
  journal={SIAM Journal on Applied Mathematics},
  volume={58},
  pages={1072--1093},
  year={1998}
}

@article{beysens11,
  title={Heat can cool near-critical fluids},
  author={Beysens, Daniel and Fr{\"o}hlich, Thomas and Garrabos, Yves},
  journal={Physical Review E - Statistical, Nonlinear, and Soft Matter Physics},
  volume={84},
  pages={051201},
  year={2011}
}

@article{zappoli03,
  title={Near-critical fluid hydrodynamics},
  author={Zappoli, Bernard},
  journal={Comptes Rendus M{\'e}canique},
  volume={331},
  pages={713--726},
  year={2003}
}

@article{beysens06,
  title={Vibrations in space as an artificial gravity?},
  author={Beysens, Daniel},
  journal={Europhysics News},
  volume={37},
  pages={22--25},
  year={2006}
}

@inproceedings{colombani00,
  title={Microgravity and earth thermal diffusion in liquids holographic visualization of convection},
  author={Colombani, Jean and Bert, Jacques},
  booktitle={AIP Conference Proceedings},
  volume={504},
  pages={861--865},
  year={2000},
  organization={American Institute of Physics}
}

@article{rodriguez20,
  title={Zero-gravity thermal convection in granular gases},
  author={Rodrigu{\'e}z-Rivas, A and L{\'o}pez-Casta{\~n}o, Miguel A and Vega Reyes, Francisco},
  journal={Physical Review E  - Statistical, Nonlinear, and Soft Matter Physics},
  volume={102},
  pages={010901},
  year={2020}
}

@article{kostoglou11,
  title={Heat transfer from small objects in microgravity: experiments and analysis},
  author={Kostoglou, Margaritis and Evgenidis, Sotiris P and Zacharias, Konstantinos A and Karapantsios, Thodoris D},
  journal={International Journal of Heat and Mass Transfer},
  volume={54},
  pages={3323--3333},
  year={2011}
}

@article{swift,
  title={Thermoacoustic engines},
  author={Swift, Gregory W},
  journal={the Journal of the Acoustical Society of America},
  volume={84},
  pages={1145--1180},
  year={1988}
}

@article{bar07,
  title={Thermal management of high heat flux nanoelectronic chips},
  author={Bar-Cohen, Avram and Wang, Peng and Rahim, Emil},
  journal={Microgravity Science and Technology},
  volume={19},
  pages={48--52},
  year={2007}
}

@inproceedings{broyan10,
  title={International space station crew quarters ventilation and acoustic design implementation},
  author={Broyan, James and Welsh, David and Cady, Scott},
  booktitle={40th International Conference on Environmental Systems},
  pages={6018},
  year={2010}
}

@article{anzini22,
  title={Fluid flow at interfaces driven by thermal gradients},
  author={Anzini, Pietro and Filiberti, Zeno and Parola, Alberto},
  journal={Physical Review E - Statistical, Nonlinear, and Soft Matter Physics},
  volume={106},
  pages={024116},
  year={2022}
}

@article{bregulla16,
  title={Thermo-osmotic flow in thin films},
  author={Bregulla, Andreas P and W{\"u}rger, Alois and G{\"u}nther, Katrin and Mertig, Michael and Cichos, Frank},
  journal={Physical Review Letters},
  volume={116},
  pages={188303},
  year={2016}
}

\end{document}